\documentclass[journal]{IEEEtran}
\usepackage[T1]{fontenc}
\usepackage[utf8]{inputenc}
\usepackage{graphicx}
\usepackage{booktabs,multirow}
\usepackage{amsmath,amssymb,amsthm}
\usepackage{cite}
\usepackage[caption=false,font=footnotesize]{subfig}
\usepackage{xcolor}
\usepackage{tcolorbox}
\usepackage[ruled,vlined,linesnumbered]{algorithm2e}
\usepackage{algpseudocode}
\usepackage{ragged2e}
\usepackage{xurl}
\usepackage{microtype}
\usepackage[hidelinks]{hyperref}
\graphicspath{{figs/}{../figs/}{trans/figs/}}

\title{Signature Knowledge Shielding: Protecting Antivirus Detection Knowledge through Robust Embedding Matching}
\author{Jieshuai~Yang,
    Zhi~Wang,
    Yan~Jia,
    Zhenhua~Wu,
    Jianfei~Tang,
    Chenbin~Su,
    Jingwei~Ye,
    Jianwen~Tian,
    and Wanpeng~Li%
    \thanks{(Corresponding author: Zhi~Wang.)}%
    \thanks{Jieshuai~Yang, Zhi~Wang, Yan~Jia, Zhenhua~Wu,
    Jianfei~Tang, Chenbin~Su, and Jingwei~Ye are with the
    College of Cryptology and Cyber Science, Nankai University,
    Tianjin 300350, China
    (e-mail: yjs@mail.nankai.edu.cn;
    zwang@nankai.edu.cn;
    jiay@nankai.edu.cn;
    2120250720@mail.nankai.edu.cn;
    kid519388073@mail.nankai.edu.cn;
    champion.su@mail.nankai.edu.cn;
    jwye@mail.nankai.edu.cn).}%
    \thanks{Jianwen~Tian is with the School of Computing and
    Information Systems, Singapore Management University,
    80 Stamford Road, Singapore 178902
    (e-mail: jwtian@smu.edu.sg).}%
    \thanks{Wanpeng~Li is with the School of Computer Science
    and Informatics, University of Liverpool, Liverpool, UK
    (e-mail: wanpeng.li@liverpool.ac.uk).}%
}
\begin{document}
\maketitle

\begin{abstract}
Antivirus (AV) signatures encode valuable detection knowledge accumulated through expert analysis, yet repeated black-box queries can expose this knowledge for unauthorized reuse and targeted evasion. We propose Signature Knowledge Shielding (SKS), a design paradigm that protects existing AV rules, termed Protected Signatures (PS), through auxiliary Shielding Signatures (SS). SKS uses AV-supervised importance ranking to discover candidate byte blocks, filters and selects them through negative-sample calibration, and combines SS matches with original AV verdicts through logical OR. SS matching introduces additional positive responses and verdict changes during probing, making detection feedback less directly attributable to PS and increasing the difficulty of localizing signature-related content, while preserving original positive verdicts. Within SKS, we integrate five individually tuned non-learning matching methods and further propose Robust Embedding Matching (REM), a lightweight learned matcher trained with random context and byte perturbations. REM provides stronger and more consistent protection against content modifications and scanning-alignment changes. We develop and open-source a ClamAV-based prototype and evaluate it against five probing methods. With an observed additional test false-positive rate below 0.1\%, REM raises the mean query count of state-of-the-art probing to as much as $282.57\times$ the unprotected baseline, while introducing only 0.67\% GPU scanning overhead. These results demonstrate that SKS can substantially increase signature-extraction costs with limited impact on detection utility and scanning efficiency.
\end{abstract}


\section{Introduction}
\label{sec:introduction}

Signature-based detection remains a core component of modern antivirus (AV) systems~\cite{botacin2022antiviruses}. Carefully curated signatures provide explicit, reproducible matching conditions for detecting known threats with low false-positive rates. These rules distill expert analysis into reusable detection knowledge, making proprietary signature collections valuable intellectual assets. However, keeping the database confidential does not eliminate knowledge leakage through the scanner's observable behavior.

An adversary can repeatedly modify a detected file and observe binary scanning verdicts to infer detection-relevant content. Prior work has investigated signature inference through such feedback~\cite{hamlen2009interface,wressnegger2017inferring}. More recently, AVHunter demonstrates that black-box queries can reveal byte-level decision-critical regions for reproducing AV behavior and guiding targeted evasion~\cite{yang2026breakingblackboxbytelevel}, complementing findings on malware-classifier and antivirus extraction~\cite{rigaki2023stealing}. Such knowledge leakage creates opportunities for unauthorized reuse of proprietary detection expertise. Bi-Iocane further shows that manipulating AV-sensitive bytes can corrupt AV-generated labels and poison downstream learning-based detectors, exposing risks in the AV-to-ML labeling supply chain~\cite{yang2026weaponizinggroundtruthdata}. Protecting signature knowledge is therefore important not only for safeguarding intellectual assets, but also for resisting evasion and preserving the integrity of downstream detection pipelines.

Existing approaches address only part of this problem. Concealing the signature database restricts direct access but does not eliminate knowledge leakage through scanning feedback~\cite{wressnegger2017inferring}. Server-enforced query controls can increase extraction costs for remote services~\cite{dziedzic2022increasing}, but their effectiveness cannot be assumed when adversaries control the local scanning environment. Filiol proposes combinatorial subpatterns and instance-specific detection functions to resist black-box analysis~\cite{filiol2006patterns}. However, this approach requires reorganizing detection patterns and changing their decision logic, rather than directly protecting existing rules in their original form. A key question therefore remains how AV providers can protect their accumulated signature rules while preserving the detection knowledge and matching mechanisms already in place.

We propose \emph{Signature Knowledge Shielding (SKS)}, a design paradigm that introduces \emph{Shielding Signatures (SS)} to protect existing rules, termed \emph{Protected Signatures (PS)}. While preserving the original detection mechanism, SKS uses SS matching to introduce additional positive responses and irregular verdict changes during probing. These responses make detection feedback harder to attribute to PS, increasing the difficulty of localizing signature-related content. Realizing this objective presents two challenges.

\textbf{Challenge \#1: Increasing signature-probing complexity.} An additional exact pattern may become another extraction target, leaving the original rules exposed once removed. Effective shielding must complicate probing feedback to substantially increase the queries required to localize protected signatures, while remaining effective against content modifications and scanning-alignment changes that adversaries may exploit.

\textbf{Challenge \#2: Preserving detection reliability.} Introducing SS can strengthen protection, but may also cause previously negative inputs to be flagged. AV systems demand highly reliable detection. Shielding must therefore preserve original detections while strictly limiting additional false positives, without imposing excessive scanning or storage costs.

Within SKS, we integrate five individually tuned non-learning matching methods and further introduce \emph{Robust Embedding Matching (REM)}, a lightweight learned matcher. The REM-based pipeline comprises four stages, namely important-block discovery, matching-model training, SS library construction, and online scanning. First, an AV-supervised model identifies and ranks important byte blocks to form candidate pools. To address \textbf{Challenge~\#1}, REM learns to match reference blocks under random context and byte perturbations. It combines tolerance to content and alignment changes with fine-grained, irregular detection responses during probing, making PS-related content harder to isolate.

To address \textbf{Challenge~\#2}, SKS uses negative samples from the training and validation sets to filter false-positive-prone candidates under the calibrated scanning configuration. Guided by importance rankings, construction proceeds for at most $R$ rounds and retains up to $k$ accepted blocks per source sample before global deduplication. During online scanning, SKS combines SS matches with the original AV verdict through logical OR, preserving every original positive verdict while introducing auxiliary matches that complicate signature probing. The objective is to protect existing signature knowledge rather than extend detection to unknown malware.

We develop and open-source a ClamAV-based SKS prototype and evaluate it against five probing strategies. Using a configuration with an observed additional test false-positive rate of 0.0891\%, REM raises the mean query count of state-of-the-art probing to as much as $282.57\times$ the unprotected baseline and retains a $272.41\times$ query multiplier under random scan-grid shifts. This configuration provides SS coverage for 95.73\% of training positives, with only 0.67\% GPU scanning overhead and a 7.92\% storage increase relative to the original ClamAV signature database. These results demonstrate protection of signature knowledge through auxiliary matching while preserving original detections and keeping additional false positives and deployment overhead manageable.

Our contributions are as follows:
\begin{itemize}
    \item \textbf{A design paradigm for shielding signature knowledge.} We propose SKS, which substantially increases signature-probing costs through auxiliary SS matching while preserving original AV positive verdicts and limiting additional false positives.
    \item \textbf{A learning-based method for SS matching.} We introduce REM, a lightweight matcher trained with random context and byte perturbations, providing stronger and more consistent protection under scanning-alignment changes than the evaluated non-learning alternatives.
    \item \textbf{An open-source ClamAV-based prototype.} We develop and open-source an SKS prototype integrating REM and five tuned non-learning matchers, supporting reproducible evaluation of signature protection, detection utility, and scanning and storage overhead.
\end{itemize}













\begin{figure*}[t]\centering
	\centering
	\includegraphics[width=1.0\linewidth]{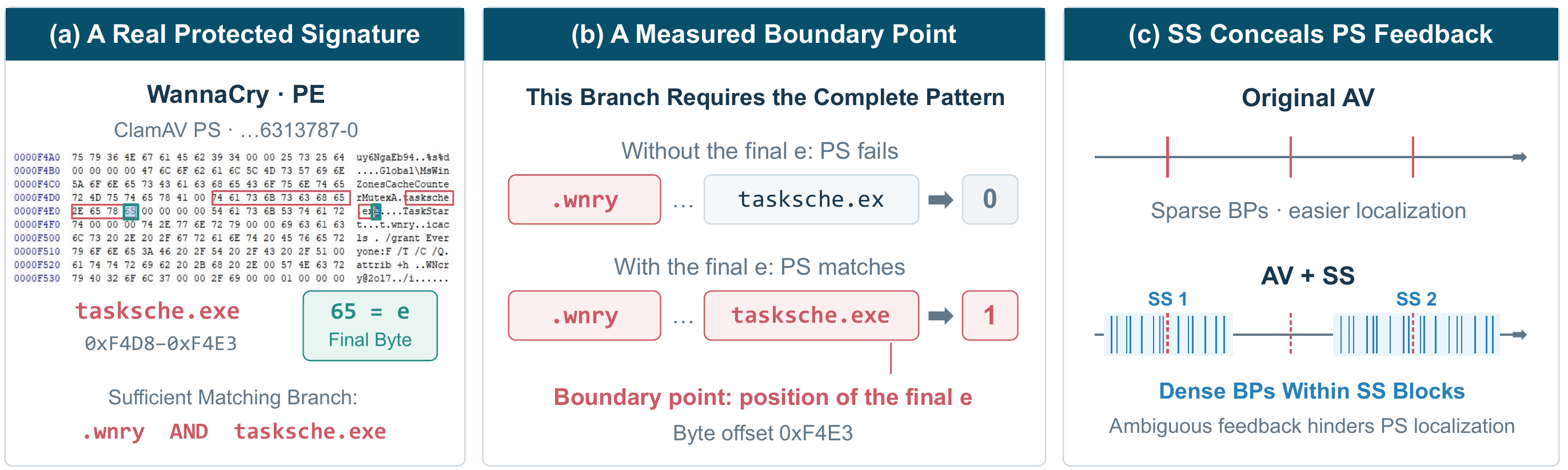}
    \caption{PS, BP, and SS illustrated with a real WannaCry sample. (a) The strings .wnry and tasksche.exe jointly satisfy a matching branch of a ClamAV signature (PS). (b) Without the final “e,” this branch fails to match, illustrating a BP at that byte position. (c) SS protection is shown schematically: dense BPs within SS blocks obscure PS-related feedback and complicate signature localization.}

	\label{fig:ps_ss_boundaries}
\end{figure*}

\section{Preliminaries}
\label{sec:preliminaries}

\subsection{Protected Signatures and Shielding Signatures}
\label{sec:protected_signatures}

Signature-based matching provides explicit conditions and efficient detection of known threats~\cite{botacin2022antiviruses}. False positives and performance degradation under distribution shifts limit reliance on ML models as the primary defense, so signature matching remains central to modern AV systems, with learning-based methods providing complementary protection~\cite{botacin2022antiviruses,cavallaro2023machine}. When learning-based detection is not triggered and verdicts depend on a few localized byte regions, iterative probing may expose detection knowledge. AVHunter, DefenderCheck, HDD, and its ProbDD- and CDD-based variants demonstrate query-based localization of detection-relevant content~\cite{yang2026breakingblackboxbytelevel,defendercheck,misherghi2006hdd,wang2021probdd,zhang2025cdd}, enabling unauthorized knowledge reuse, targeted evasion, and poisoning of AV-based labeling supply chains~\cite{rigaki2023stealing,yang2026breakingblackboxbytelevel,yang2026weaponizinggroundtruthdata}.

We call existing AV signatures selected for protection \emph{Protected Signatures} (PS). Common types include file and section hashes, wildcard byte patterns, and logical signatures combining multiple conditions~\cite{clamav_signatures}. These rules encode knowledge accumulated through expert analysis and maintenance. We focus on PS depending on localized byte regions, which are often sparse within a file and can be isolated through iterative probing. Figure~\ref{fig:ps_ss_boundaries}(a) illustrates such content in a real WannaCry sample. Global signatures, such as whole-file hashes, lack this spatial property.

\emph{Shielding Signatures} (SS) are constructed from selected byte blocks and stored separately as matcher-specific representations, including REM embeddings and non-learning fingerprints or digests. SKS uses REM and integrates five tuned non-learning matchers as alternatives and baselines. SS neither modify PS nor insert bytes into scanned files. Their matches introduce additional positive responses that obscure PS-related feedback and increase signature-localization costs.
\subsection{Boundary Points}
\label{sec:boundary_points}

A \emph{boundary point} (BP) is a byte position associated with a change in the detection verdict under a specified modification and probing procedure. Such a position may be identified by advancing a masking boundary or by removing a byte from a reduced input. BPs therefore depend on the modification procedure and the current input state; different probing methods may locate different BPs in the same sample. Figure~\ref{fig:ps_ss_boundaries}(b) illustrates a BP through the loss of a byte required by the displayed PS matching branch. An adversary observing only binary verdicts can iteratively locate such positions and examine their surrounding bytes to infer detection-relevant content. AVHunter validates this connection against ClamAV's signature database: among samples with recoverable ground-truth matching regions, 97.39\% exhibit overlap between these regions and the probed critical bytes~\cite{yang2026breakingblackboxbytelevel}. Beyond ClamAV, its experiments on commercial AV products show that modifying inferred boundary regions can enable evasion, while inserting such regions into benign files can trigger detections, providing evidence of their association with AV decisions~\cite{yang2026breakingblackboxbytelevel}. These findings support using BPs and surrounding context to construct and validate candidate byte-pattern rules.

\subsection{Threat Model}
\label{sec:adversary_capabilities}

We focus on protecting AV signatures used for static detection of Windows PE files. We consider an adversary with black-box query access to an AV scanner protected by SKS. The adversary can obtain detected PE samples, submit modified inputs, and observe scanning responses. All reported detections are treated as positive outcomes, while no detection is treated as a negative outcome.

The adversary can adaptively replace, mask, truncate, insert, or delete bytes based on previous responses and retain query histories to isolate detection-relevant content. Queried inputs need not remain valid PE files. We allow knowledge of the SKS design and matching procedure, but assume no direct access to the PS or SS libraries and no ability to modify the scanner's decision logic.

The adversary aims to localize PS-related content and recover static detection knowledge. We evaluate its progress under a bounded query budget to approximate time and resource constraints. Our scope concerns static signatures rather than behavioral detection during program execution.

\subsection{Motivation and Protection Objective}
\label{sec:motivation_objective}

Under the threat model described in Section~\ref{sec:adversary_capabilities}, existing probing methods progressively localize and modify detection-triggering regions in PE samples, enabling efficient knowledge extraction when relatively few BPs are encountered. We observe that statistical and learned approximate matching rules can produce numerous BPs during input modification. This motivates constructing dedicated SS matched using these techniques: their matches introduce additional positive responses and boundaries, making PS-related feedback harder to isolate, as illustrated schematically in Figure~\ref{fig:ps_ss_boundaries}(c). Building on this idea, we propose SKS, a design paradigm that combines candidate discovery, calibrated SS library construction, and matching techniques that generate numerous BPs to protect existing static detection signatures.

Our objective is to increase the query costs of localizing and extracting PS in the Windows PE setting, hindering large-scale knowledge extraction under practical time and resource constraints. Protection should preserve existing detections while limiting additional false positives. We also evaluate scanning latency and storage overhead. SKS aims to make extraction more costly, without assuming that all extraction attempts can be prevented.

\begin{figure*}[t]
\centering
\includegraphics[width=\linewidth]{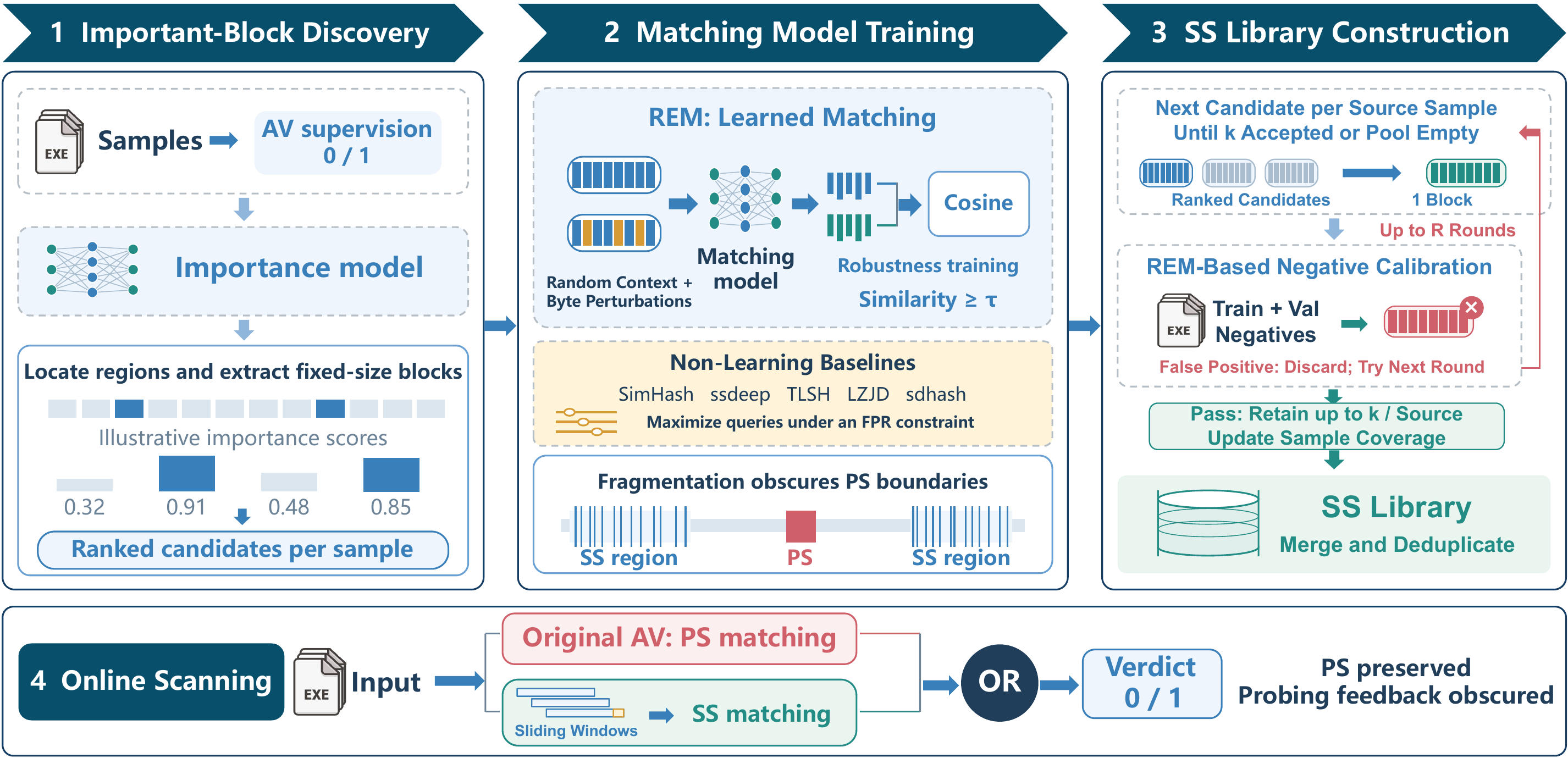}
\caption{SKS overview. (1) AV-supervised important-block discovery. (2) REM training with context and byte perturbations; five tuned non-learning matchers provide alternatives. (3) Negative calibration and source-match checks retain up to $k$ ranked blocks per sample over at most $R$ rounds. (4) Sliding-window SS matches are OR-combined with AV verdicts, preserving detections while complicating PS localization.}
\label{fig:sks_framework}
\end{figure*}

\section{Approach}
\label{sec:approach}

SKS protects existing PS by introducing SS whose matching responses obscure PS-related probing feedback. Figure~\ref{fig:sks_framework} presents its four stages. Important-block discovery produces ranked candidate pools under AV supervision. Matching model training uses these blocks to learn REM, which tolerates surrounding context and content modifications. SS library construction then applies REM to filter candidates through multiple rounds of selection, accumulating suitable blocks from each sample. Online scanning reuses the trained matcher and the resulting SS library, combining SS matches with the original AV verdict. By introducing additional positive responses and fragmented, irregular boundaries, SKS makes PS-related content harder to isolate and increases signature-localization queries. The following stages describe the REM-based implementation; non-learning alternatives retain their native representations and acceptance rules.

\subsection{Important-Block Discovery}
\label{sec:block_discovery}

SKS first identifies byte blocks associated with the target AV's detection behavior. We train an importance model using binary labels supplied by the AV, guiding candidate discovery without requiring access to the underlying PS.

The importance model is a shallow multilayer perceptron. It maps an 80-dimensional representation of byte frequencies and hashed byte bigrams through layers of 128 and 64 units, followed by a linear scoring head. Following a multiple-instance learning formulation~\cite{carbonneau2018multiple,ilse2018attention}, we treat each file as a bag of byte blocks and use max pooling to aggregate block scores into a file-level prediction supervised by AV labels. The resulting block scores guide candidate ranking. We choose this lightweight architecture because this stage primarily requires an informative ordering of candidate blocks under file-level supervision. Its low computational cost supports scoring large candidate pools and retaining alternatives for subsequent filtering.

For each AV-positive training sample $x$, the model ranks candidate blocks of length $L$. Basic filtering removes unsuitable blocks, and identical candidates within the sample are consolidated. The resulting pool $\mathcal{C}_x$ preserves importance ordering and retains alternatives for later selection.

These pools connect discovery to the next two stages. Their blocks provide reference content for REM training, while their rankings guide SS selection after the matcher is trained. Importance alone does not establish suitability: a highly ranked block may also match negative content. Such candidates are identified using REM during library construction.

\subsection{Matching Model Training}
\label{sec:matching_training}
\label{sec:ss_matching}

REM learns how to match candidate content under modification. Its encoder $f_\theta$ is trained separately from the importance model: importance ranking determines which blocks to consider, whereas REM determines whether input content sufficiently resembles a reference block.

REM uses a lightweight byte-sequence encoder, drawing on byte-level convolutional modeling~\cite{raff2017malware,raff2021extreme} and attention-based pooling~\cite{ilse2018attention,lee2019settransformer}. It comprises 8-dimensional byte embeddings, a 32-channel local convolution with kernel size 8 and stride 1, one residual feed-forward module, and attention pooling with two learned queries. A projection head produces normalized 128-dimensional embeddings. Local convolution captures ordered byte patterns, while pooling and random-context training reduce dependence on absolute positions. Unlike fixed statistical representations, REM can learn tolerance to positional changes and content modifications. This compact architecture supports protective matching while limiting additional scanning overhead.

We draw reference blocks from the training candidate pools and pair them with perturbed views. Random prefix and suffix context totaling $\Delta$ bytes places an $L$-byte reference block at varying positions within an $L+\Delta$-byte view. Random byte replacements additionally simulate content modifications. Following the principle of representation consistency across augmented views~\cite{chen2020simclr}, training combines contrastive cross-entropy with squared hinge penalties enforcing positive-similarity floors, a positive--hard-negative margin, and a negative-similarity ceiling. A variance regularizer discourages embedding collapse. Negatives include mined hard windows, randomly sampled negative windows, and padded negative views from AV-negative training files.

REM compares an input block $b$ and reference block $s$ using cosine similarity:
\begin{equation}
\operatorname{sim}_{\theta}(b,s)=
\frac{f_\theta(b)^\top f_\theta(s)}
{\lVert f_\theta(b)\rVert_2
 \lVert f_\theta(s)\rVert_2}.
\label{eq:rem_similarity}
\end{equation}
The similarity in Equation~\ref{eq:rem_similarity} is converted into a block-level decision using a unified threshold $\tau$:
\begin{equation}
A_{\mathrm{REM}}(b,s;\Gamma)=
\mathbf{1}\!\left[
\operatorname{sim}_{\theta}(b,s)\geq\tau
\right],
\label{eq:rem_block_match}
\end{equation}
where $\Gamma$ contains the trained encoder and matching configuration. The acceptance rule in Equation~\ref{eq:rem_block_match} is fixed before SS selection and reused during online scanning, ensuring that candidate filtering evaluates the same matching behavior used at deployment.

SKS also integrates five non-learning approximate matchers based on statistical features and hand-designed matching rules, including SimHash~\cite{charikar2002similarity}, ssdeep~\cite{kornblum2006identifying}, TLSH~\cite{oliver2013tlsh}, LZJD~\cite{raff2018lzjd}, and sdhash~\cite{roussev2010fingerprinting}. Each retains its native representation and comparison rule and is tuned for PS protection under constraints on additional false positives, serving as both an alternative matcher and a comparison baseline. Exact byte matching serves as a control.

\subsection{SS Library Construction}
\label{sec:ss_construction}

This stage takes the ranked candidate pools and the trained REM matcher as inputs and constructs the SS library $S$. Its objective is to provide broad sample coverage and retain multiple suitable protective blocks while limiting additional false positives. Coverage is the proportion of AV-positive training samples matched by at least one SS under the online scanning configuration.

Starting from an empty library, construction proceeds for at most $R$ rounds, retaining up to $k$ accepted blocks per source sample. In each round, we select the highest-ranked, untried candidate from every sample that has remaining candidates and fewer than $k$ accepted blocks. Samples already covered by the current library remain eligible for selection. Thus, $R$ limits the number of candidates examined per sample, whereas $k$ limits the number retained before global deduplication.

Each selected block is evaluated as a reference SS against negative files from the training and validation sets. This check uses REM with the calibrated configuration, including the similarity threshold, window extraction, and padding policy. A block is discarded if it matches any calibration negative. Otherwise, it is accepted only if it matches a scanning window of its source file.

Accepted blocks are added to $S$, and coverage is updated. A rejected candidate does not count toward the $k$-block limit; the sample advances to its next highest-ranked candidate in the following round. Rejected blocks are never reintroduced. Selection continues for each sample until $k$ blocks are accepted, its candidate pool is exhausted, or the $R$-round limit is reached. Continuing selection for covered samples provides additional protective matches, reducing dependence on a single SS.

Finally, identical accepted blocks are deduplicated globally, and their normalized REM embeddings are stored as the deployed SS library. Raw reference byte blocks are not included in the reported deployment storage. The resulting Top-$k$ library therefore retains up to $k$ eligible blocks per source sample before deduplication, rather than necessarily retaining its first $k$ candidates. The encoder and calibration configuration remain fixed throughout construction. The resulting library is frozen before evaluation, and test samples are not used for candidate selection or filtering.

\subsection{Online Scanning}
\label{sec:online_scanning}

Online scanning applies the frozen SS library and calibrated matching configuration. We describe REM-based scanning below; alternative matchers use native representations and comparison rules.

For an input $x$ of length $n$, windows start at offset zero with stride $\Delta$ and length $\ell=L+\Delta$. Following the configuration analysis in Section~\ref{sec:reference_block_stride}, we set $L=1{,}024$ bytes and $\Delta=128$ bytes, yielding $\ell=1{,}152$-byte scanning windows. Setting $\ell=L+\Delta$ ensures that any complete $L$-byte region is fully contained in at least one scanning window despite grid misalignment. Equation~\ref{eq:input_blocks} defines the window set, including incomplete tail windows right-padded with \texttt{0x90}:
\begin{equation}
\begin{aligned}
b_i &=
\operatorname{Pad}_{\texttt{0x90}}
\bigl(x[i:\min(i+\ell,n)],\ell\bigr),\\
\mathcal{W}(x) &=
\{b_i\mid i=j\Delta<n,\ j\in\mathbb{N}_0\}.
\end{aligned}
\label{eq:input_blocks}
\end{equation}
Retaining tail windows preserves opportunities to match protective content remaining after truncation. This robustness design in SKS mitigates the risk that truncation-based probing, such as DefenderCheck, leaves important content in incomplete tail windows that would otherwise be skipped. Padding affects only SS matching and leaves the original AV input unchanged.

REM encodes the windows in Equation~\ref{eq:input_blocks} in batches and compares their normalized embeddings with cached SS embeddings. Local convolutional features can be reused across overlapping windows to reduce repeated computation. Equation~\ref{eq:file_ss_match} defines a file-level SS match whenever at least one window satisfies the block-level acceptance rule in Equation~\ref{eq:rem_block_match}:
\begin{equation}
M(x,s;\Gamma)=
\mathbf{1}\!\left[
\exists b\in\mathcal{W}(x):
A_{\mathrm{REM}}(b,s;\Gamma)=1
\right].
\label{eq:file_ss_match}
\end{equation}
SS matching stops once any match is found; an empty window set or SS library produces no protective match. The original AV evaluates $x$ using its existing PS, and Equation~\ref{eq:sks_verdict} combines its verdict with SS matches through logical OR:
\begin{equation}
D_{\mathrm{SKS}}(x)=
D_{\mathrm{AV}}(x)
\lor
\bigvee_{s\in S}M(x,s;\Gamma).
\label{eq:sks_verdict}
\end{equation}
The composition in Equation~\ref{eq:sks_verdict} preserves every original positive verdict, while negative-sample calibration limits additional false positives.

SS matches make detection feedback harder to attribute to PS. During modification, an SS may maintain a positive verdict independently of PS, while learned similarities can repeatedly cross the matching threshold and introduce irregular BPs. These responses intermingle with PS-related boundaries, requiring more probing to distinguish origins and isolate PS-related content.

\section{Evaluation}
\label{sec:evaluation}

We evaluate our ClamAV-based SKS prototype's protection of signature knowledge against black-box probing. We examine REM and five individually tuned non-learning matchers across five probing strategies, without and with scan-grid shifts. We investigate protective coverage, additional false positives, scanning and matching configurations, robustness design, distinct BP counts, and deployment overhead, including how design choices affect SS library storage. Our evaluation addresses four research questions:

\begin{itemize}
    \item \textbf{RQ1: Signature Protection Effectiveness.} To what extent does SKS increase probing queries across different probing strategies, and how does REM compare with tuned non-learning matchers without and with scan-grid shifts?

    \item \textbf{RQ2: Detection Utility and Additional False Positives.} How effectively does SKS cover known detected samples while preserving original positive verdicts and limiting additional false positives on test negatives?

    \item \textbf{RQ3: Design Choices and Boundary Point Counts.} How do reference-block length and scanning stride balance protection, coverage, and overhead? How do matching thresholds and block-retention limits affect additional false positives? Does robustness design increase probing costs, and how does SS matching affect distinct BP counts within budget?

    \item \textbf{RQ4: Runtime and Storage Overhead.} What additional scanning and storage overhead does SKS introduce across matchers, and how do vector precision and block selection affect SS library storage while maintaining coverage and limiting additional false positives?
\end{itemize}

\subsection{Evaluation Setup}


\subsubsection{Datasets}

Our dataset comprises 176,892 PE files primarily from SOREL-20M, BODMAS, and VirusShare~\cite{harang2020sorel,yang2021bodmas,virusshare}. We demonstrate signature protection using ClamAV and assign samples its binary labels, as our objective is to protect existing AV signature knowledge rather than learn more generalizable malware-detection knowledge. Motivated by findings on temporal evaluation bias and concept drift~\cite{pendlebury2019tesseract,arp2022dos,chen2023continuous}, we sort each label class by PE COFF timestamp and chronologically split it into approximately 80\% training, 10\% validation, and 10\% test data. Table~\ref{tab:dataset} reports partition sizes. Training samples support importance-model training and candidate block extraction. Training and validation negatives calibrate the SS library and matching parameters to constrain additional false positives, which the test set measures relative to ClamAV verdicts.


\begin{table}[t]
\centering
\caption{Dataset partitions with ClamAV-derived labels.}
\label{tab:dataset}
\small
\setlength{\tabcolsep}{4pt}
\renewcommand{\arraystretch}{1.15}
\begin{tabular}{lrrrr}
\toprule[1.1pt]
Label & Training & Validation & Test & Total \\
\hline
Positive & 69,474  & 8,684  & 8,682  & 86,840  \\
Negative & 73,470  & 8,723  & 7,859  & 90,052  \\
Total    & 142,944 & 17,407 & 16,541 & 176,892 \\
\bottomrule[1.1pt]
\end{tabular}
\end{table}



\subsubsection{SS Matching Methods and Calibration}
SKS supports REM and non-learning SS matchers based on statistical features and heuristic similarity measures. SimHash compares binary fingerprints using Hamming distance~\cite{charikar2002similarity}; ssdeep compares context-triggered piecewise hashes~\cite{kornblum2006identifying}; TLSH measures distances between digests of local byte statistics~\cite{oliver2013tlsh}; LZJD estimates Jaccard similarity between Lempel--Ziv substring sets~\cite{raff2018lzjd}; and sdhash compares Bloom-filter digests constructed from statistically discriminative features~\cite{roussev2010fingerprinting}. Exact requires identical byte content and serves as a basic control. REM instead compares normalized embeddings produced by an encoder trained with random context and byte perturbations.

We tune the non-learning matchers for PS protection with a target additional false-positive rate below 0.2\%, calibrating their acceptance rules against training and validation negatives and reporting the resulting test-set false-positive rates separately. The selected SimHash configuration uses 256-bit fingerprints, a maximum matching radius of 80, and a calibration margin of 16 relative to the nearest negative fingerprint. The other configurations use ssdeep scores of at least 95, TLSH distances of at most 15, LZJD similarity of at least 0.92, and sdhash scores of at least 92. These methods serve as comparison baselines and potential alternatives for lightweight SKS deployments.


\subsubsection{Probing Methods and Protocol}
We evaluate five probing methods: AVHunter, which locates verdict-changing byte positions through boundary search~\cite{yang2026breakingblackboxbytelevel}; DefenderCheck~\cite{defendercheck}, which recursively splits inputs to identify detection-triggering bytes; HDD~\cite{misherghi2006hdd}, which hierarchically reduces inputs while preserving a positive verdict; and HDD variants incorporating probabilistic subset selection from ProbDD~\cite{wang2021probdd} or counter-based subset selection from CDD~\cite{zhang2025cdd}. For all methods, we apply a common strategy to handle repeated BPs: when the same BP is repeatedly reported, we mask surrounding bytes and progressively expand the masked region until probing advances to another BP, avoiding infinite loops at the same position. Probing stops at the query budget or when the target AV labels the cumulatively masked input negative. We emphasize defending against methods requiring fewer queries against the unprotected scanner, as their lower extraction cost makes them more attractive choices for resource-constrained adversaries.

\begin{table*}[t]
\centering
\renewcommand{\arraystretch}{1.1}
\setlength{\tabcolsep}{5pt}
\caption{Mean total probing queries without and with random scan-grid shifts.}
\label{tab:rq1_queries}

\resizebox{0.9\linewidth}{!}{%
\begin{tabular}{cccccccc}
\toprule[1.1pt]
Setting & Method / Configuration & AVHunter & DefenderCheck & HDD & HDD+ProbDD & HDD+CDD & Average \\
\midrule
\multirow{10}{*}{\shortstack{Without\\scan-grid shifts}}
& Unprotected & 54.34 & 357.54 & 4,480.66 & 1,069.82 & 1,018.44 & 1,396.16 \\
& Exact & 332.08 & 1,491.18 & 27,061.82 & 18,578.50 & 12,685.68 & 12,029.85 \\
& SimHash~\cite{charikar2002similarity} & 13,687.54 & 18,636.96 & \textbf{45,000.20} & \textbf{34,939.68} & 33,682.48 & 29,189.37 \\
& ssdeep~\cite{kornblum2006identifying} & 264.48 & 1,070.16 & 29,856.14 & 12,021.88 & 11,239.44 & 10,890.42 \\
& TLSH~\cite{oliver2013tlsh} & 58.52 & 364.60 & 5,293.06 & 1,196.68 & 1,103.98 & 1,603.37 \\
& LZJD~\cite{raff2018lzjd} & 209.94 & 478.20 & 9,404.18 & 3,983.76 & 3,427.02 & 3,500.62 \\
& sdhash~\cite{roussev2010fingerprinting} & 749.66 & 1,923.26 & 31,084.16 & 11,900.90 & 10,319.14 & 11,195.42 \\
\cmidrule(lr){2-8}
& REM ($\tau=0.95$, Top3) & 7,358.02 & 9,464.84 & 41,013.16 & 24,060.94 & 25,688.14 & 21,517.02 \\
& REM ($\tau=0.91$, Top3) & 15,354.86 & 16,686.44 & 41,609.24 & 28,788.36 & 29,551.46 & 26,398.07 \\
& REM ($\tau=0.90$, Top7) & \textbf{20,467.02} & \textbf{20,956.72} & 43,000.28 & 33,741.54 & \textbf{34,526.58} & \textbf{30,538.43} \\
\midrule
\multirow{10}{*}{\shortstack{With random\\scan-grid shifts}}
& Unprotected & 54.34 & 357.54 & 4,480.66 & 1,069.82 & 1,018.44 & 1,396.16 \\
& Exact & 332.08 & 1,491.18 & 27,061.82 & 18,578.50 & 12,685.68 & 12,029.85 \\
& SimHash~\cite{charikar2002similarity} & 2,038.76 & 2,884.10 & 15,589.48 & 7,751.56 & 7,630.48 & 7,178.88 \\
& ssdeep~\cite{kornblum2006identifying} & 98.62 & 514.06 & 16,250.48 & 3,296.30 & 3,714.30 & 4,774.75 \\
& TLSH~\cite{oliver2013tlsh} & 57.66 & 364.30 & 5,737.64 & 1,297.70 & 1,199.16 & 1,731.29 \\
& LZJD~\cite{raff2018lzjd} & 121.82 & 396.00 & 6,185.20 & 2,593.82 & 2,267.64 & 2,312.90 \\
& sdhash~\cite{roussev2010fingerprinting} & 675.24 & 1,560.72 & 29,587.42 & 8,133.20 & 7,174.08 & 9,426.13 \\
\cmidrule(lr){2-8}
& REM ($\tau=0.95$, Top3) & 6,798.10 & 8,665.48 & 40,111.84 & 22,630.34 & 24,336.68 & 20,508.49 \\
& REM ($\tau=0.91$, Top3) & 14,802.56 & 15,982.94 & 41,770.28 & 28,426.94 & 29,616.06 & 26,119.76 \\
& REM ($\tau=0.90$, Top7) & \textbf{19,816.10} & \textbf{20,355.00} & \textbf{43,000.28} & \textbf{33,254.96} & \textbf{33,474.68} & \textbf{29,980.20} \\
\bottomrule[1.1pt]
\end{tabular}%
}

\smallskip
\begin{minipage}{\linewidth}
\footnotesize
Entries report mean total queries over the same 500 samples covered by all evaluated SS libraries before shifts. Each shifted sample uses a fixed random scan-grid offset of 1--1,024 bytes, shared across methods; file content is unchanged. Average is the arithmetic mean across five probing methods. Budget-reaching runs count as 50,000 queries. REM ($\tau=0.95$, Top3), REM ($\tau=0.91$, Top3), and REM ($\tau=0.90$, Top7) target additional FPR budgets of 0.05\%, 0.10\%, and 0.20\%, respectively.
\end{minipage}
\end{table*}

\subsubsection{Evaluation Metrics}
\label{sec:evaluation_metrics}

We use the following metrics to evaluate protection effectiveness, boundary behavior, detection utility, and overhead.

\begin{itemize}
    \item \textbf{Probing Queries} measures total queries ($Q_{\mathrm{tot}}$) consumed by each probing run. Runs reaching the query budget are included at the cap. Higher counts indicate greater probing costs.

    \item \textbf{Distinct Boundary Points} measures the number of distinct BPs reported within the query budget, characterizing the boundary behavior encountered during probing.

    \item \textbf{Coverage and Additional False-Positive Rate} measure the fraction of AV-positive samples matched by at least one SS and the fraction of AV-negative samples additionally flagged by SS, respectively. Additional FPR is relative to ClamAV verdicts, rather than independently verified benign ground truth.

    \item \textbf{Scanning and Storage Overhead} measures additional scanning latency relative to unmodified ClamAV and SS library storage, with relative storage increases including the shared REM model where applicable.
\end{itemize}

\subsubsection{Experimental Setup}

All experiments run on Windows 11 with an NVIDIA GeForce RTX 4080 SUPER GPU and an Intel Core i7-14700K CPU. We develop and evaluate SKS on ClamAV 1.3.1, whose open-source implementation and publicly available signature database facilitate inspection and validation, providing a reference implementation of knowledge protection for closed-source AV vendors. We freeze main.cvd, daily.cvd, and bytecode.cvd at versions 62, 27366, and 335, respectively. The importance model trains on blocks from AV-positive and AV-negative training files with file-level AV supervision for 10 epochs, retaining the checkpoint with the lowest validation loss. REM learns from reference blocks and views augmented with random context and byte perturbations, producing normalized 128-dimensional embeddings after 16 initial epochs and three 8-epoch continuation stages. We evaluate $(\tau,k)=(0.95,3)$, $(0.91,3)$, and $(0.90,7)$ for target additional FPR budgets of 0.05\%, 0.10\%, and 0.20\%, respectively, where $\tau$ is the similarity threshold and $k$ caps accepted blocks per source sample before deduplication. We set the construction-round limit to $R=k$ throughout the experiments. The three primary REM configurations share the same trained models, and SS libraries are frozen before probing evaluation.

\begin{figure*}[t]\centering
	\centering
	\includegraphics[width=1.0\linewidth]{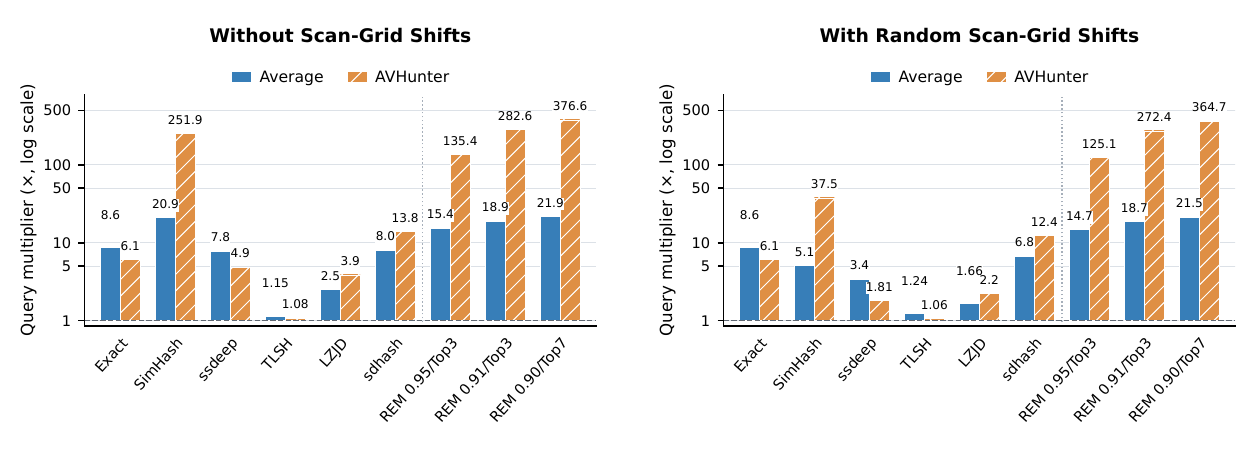}
    \caption{Probing-query multipliers relative to unprotected detection, without scan-grid shifts (left) and with random scan-grid shifts (right). Each method uses average total queries across five probing strategies and AVHunter separately. Larger values indicate higher probing costs. The vertical axis uses a logarithmic scale.}
	\label{fig:RQ1_Query_Multipliers_pair_compact}
\end{figure*}

\begin{table*}[t]
\renewcommand{\arraystretch}{1.15}
\caption{SS coverage and additional test false-positive rates of REM and non-learning matching baselines.}
\label{tab:rq2_coverage_fpr}

{%
\centering
\resizebox{0.7\linewidth}{!}{%
\begin{tabular}{ccccc}
\toprule[1.1pt]
Method / Configuration &
\shortstack{Training SS\\Coverage (\%)} &
\shortstack{Validation SS\\Coverage (\%)} &
\shortstack{Test SS\\Coverage (\%)} &
\shortstack{Additional Test\\FPR (\%)} \\
\midrule
SimHash~\cite{charikar2002similarity} & \textbf{98.52} & 5.54 & 1.83 & 0.1272 \\
ssdeep~\cite{kornblum2006identifying}  & 88.11 & 5.30 & 1.22 & 0.0254 \\
TLSH~\cite{oliver2013tlsh}    & 29.36 & 3.94 & 0.51 & 0.0000 \\
LZJD~\cite{raff2018lzjd}    & 41.95 & 1.60 & 0.63 & 0.0000 \\
sdhash~\cite{roussev2010fingerprinting}  & 85.44 & 6.11 & 1.13 & 0.0127 \\
\midrule
REM ($\tau=0.95$, Top3)
& 94.19 & 10.01 & 1.37 & 0.0254 \\
REM ($\tau=0.91$, Top3)
& 95.73 & 14.47 & 1.65 & 0.0891 \\
REM ($\tau=0.90$, Top7)
& 96.66 & \textbf{16.69} & \textbf{2.66} & 0.1909 \\
\bottomrule[1.1pt]
\end{tabular}%
}
\par
}

\smallskip
\noindent
\begin{minipage}{\linewidth}
\footnotesize
\justifying
\noindent
SS coverage is the fraction of ClamAV-positive files matching any SS across the full dataset. Additional test FPR is the fraction of ClamAV-negative test files flagged by SS. Logical OR preserves original positive verdicts. All methods use identical splits and scanning-window policies with their respective frozen SS libraries and matching configurations.
\par
\end{minipage}
\end{table*}

\subsection{RQ1: Signature Protection Effectiveness}

We evaluate signature protection using five probing methods on 500 samples jointly covered by the SS libraries of all evaluated SKS matching configurations, ensuring that every configuration provides protective matches before scan-grid shifts. Each run is capped at 50,000 queries. Given the substantial runtime of probing under protection, we choose this sample size and query budget to complete the experiments within a manageable time frame, while allowing a query volume that we consider a substantial probing burden. Table~\ref{tab:rq1_queries} reports the mean total queries without and with scan-grid shifts, including runs that reach the budget.

\paragraph{Protection without scan-grid shifts.} The upper part of Table~\ref{tab:rq1_queries} shows that REM provides the strongest overall protection: the $(\tau=0.90,k=7)$ configuration achieves the highest average query count and leads on three of the five probing methods. Non-learning matchers also provide protection, particularly SimHash, which achieves the highest query counts against HDD and HDD+ProbDD. These results show that both learned and non-learning matching can increase probing costs when scanning alignment remains unchanged.

\paragraph{Protection under scan-grid shifts.} Prior research studied functionality-preserving adversarial transformations against malware detectors, highlighting the importance of evaluating robustness to realizable input modifications~\cite{pierazzi2020problemspace,demetrio2021functionality,lucas2023adversarial,he2024malwaretotal}. An attacker aware of the matching algorithm may shift section data through header padding and corresponding PE header-field updates while preserving functionality~\cite{demetrio2021exemples}. Minor offset differences may also occur across versions of the same sample. Such changes alter the alignment between file content and scanning windows, potentially weakening alignment-sensitive SS matching. To evaluate alignment sensitivity, SS windows start at a fixed random offset of 1--1,024 bytes per sample, shared across methods, with unchanged stride. Window-based SS matching skips the preceding prefix, while PS scanning receives the full input. This simulates section-relocation-induced alignment changes without modifying PE files. The lower part of Table~\ref{tab:rq1_queries} shows that REM's advantage becomes more pronounced under shifts. Its strongest configuration leads against all five probing methods, whereas several non-learning matchers, especially SimHash and ssdeep, lose substantial protection. REM maintains protection more consistently across scanning alignments. All evaluations in RQ2--RQ4 use the same offset policy on their respective sample sets.

\paragraph{Magnitude of protection.} Figure~\ref{fig:RQ1_Query_Multipliers_pair_compact} summarizes query multipliers relative to unprotected detection. The strongest REM configuration increases average probing queries to $21.87\times$ the baseline without shifts and $21.47\times$ with shifts, with corresponding AVHunter multipliers of $376.65\times$ and $364.67\times$. Its mean query count decreases by only 1.83\% under shifts, compared with 75.41\% for SimHash (Table~\ref{tab:rq1_queries}). These results demonstrate a substantial and persistent increase in probing costs, although higher query consumption alone does not imply that signature extraction is prevented.

\begin{tcolorbox}[colback=gray!20,colframe=gray!20,boxrule=0pt,arc=0pt,left=6pt,right=6pt,top=6pt,bottom=6pt]
\textbf{Answer to RQ1.} SKS substantially increases signature-probing costs. Within the SKS framework, non-learning matchers provide some protection in non-adversarial settings where scanning alignment remains unchanged. REM achieves the strongest protection and maintains effectiveness under adversarial scan-grid shifts, providing stronger protection against probing.
\end{tcolorbox}

\begin{figure*}[t]
\centering
\includegraphics[width=\linewidth]{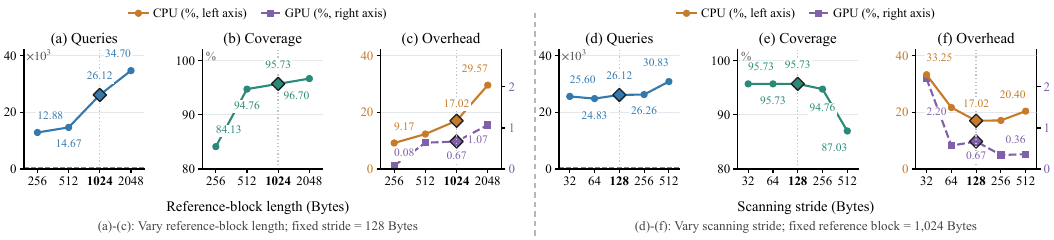}
\caption{Effects of reference-block length and scanning stride on probing cost, coverage, and overhead.}
\label{fig:rq3_block_stride_trends}
\end{figure*}

\begin{figure*}[t]\centering
	\centering
	\includegraphics[width=0.9\linewidth]{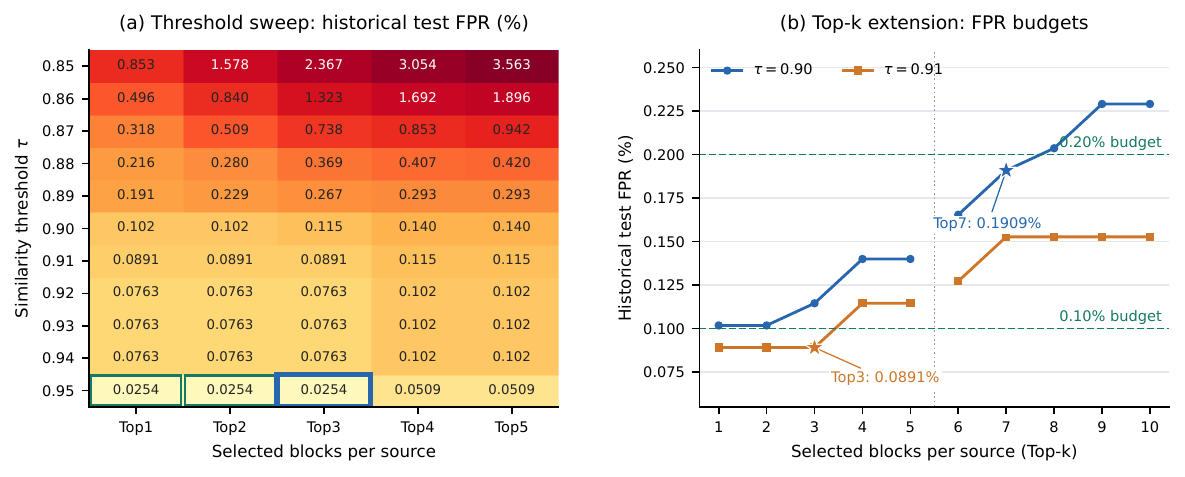}
    \caption{Matching threshold and block-selection exploration. (a) Historical test FPR across thresholds and Top-$k$ settings. (b) Extended Top-$k$ comparisons at $\tau=0.90$ and $0.91$. Highlighted configurations represent FPR budgets of 0.05\%, 0.10\%, and 0.20\%.}

	\label{fig:configuration_selection}
\end{figure*}

\subsection{RQ2: Detection Utility and Additional False Positives}

We evaluate whether SKS can protect known samples while limiting its impact on the original detector. Table~\ref{tab:rq2_coverage_fpr} reports SS coverage and additional false-positive rates. SKS aims to protect existing signature knowledge rather than extend detection to unknown malware.

\paragraph{Protection of known samples.} REM achieves 94.19--96.66\% coverage on training positives, providing protective matches for most samples used in SS construction. SimHash also achieves high training coverage, at 98.52\%. Coverage is lower on validation and test positives, but detecting unknown malware is not the objective of SS, and the original signature library may fail to detect such malware.

\paragraph{Additional false positives on unseen samples.} A key requirement is that protective matching should minimize additional positive verdicts on negative samples unseen during model training and candidate filtering. The test partition originates from the latest 10\% of timestamp-ordered samples within each label class. On 7,859 ClamAV-negative test files, the three REM configurations produce additional FPRs of only 0.0254\%, 0.0891\%, and 0.1909\%, respectively, all below their corresponding target budgets. The strictest configuration covers 94.19\% of training positives while introducing an additional test FPR of only 0.0254\%. These results show that broad protection of known samples can coexist with a very low observed additional false-positive rate on temporally later ClamAV-negative samples. Since this partition also supports configuration exploration, the rates describe the explored partition rather than independent held-out validation.

\begin{tcolorbox}[colback=gray!20,colframe=gray!20,boxrule=0pt,arc=0pt,left=6pt,right=6pt,top=6pt,bottom=6pt]
\textbf{Answer to RQ2.} SKS protects known samples while preserving the original detector's positive verdicts. REM covers most SS construction samples while maintaining low observed additional false-positive rates on temporally later ClamAV-negative samples unseen during training.
\end{tcolorbox}

\begin{table*}[t]
\centering
\caption{Average number of distinct BPs reported within a
50,000-query budget.}
\label{tab:rq3_boundary_points}
\renewcommand{\arraystretch}{1.1}
\resizebox{0.7\textwidth}{!}{%
\begin{tabular}{lccccc}
\toprule[1.1pt]
\multicolumn{1}{c}{Protection Method} &
AVHunter & DefenderCheck & HDD & HDD+ProbDD & HDD+CDD \\
\midrule
\multicolumn{1}{c}{Unprotected}
& 2.72 & 2.56 & 2.46 & 2.32 & 2.46 \\
\multicolumn{1}{c}{Exact}
& 16.34 & 15.04 & \textbf{8.74} & 13.70 & 15.68 \\
\multicolumn{1}{c}{SimHash~\cite{charikar2002similarity}}
& 97.58 & 31.80 & 4.28 & 10.50 & 12.74 \\
\multicolumn{1}{c}{ssdeep~\cite{kornblum2006identifying}}
& 4.90 & 4.34 & 3.82 & 3.96 & 4.94 \\
\multicolumn{1}{c}{TLSH~\cite{oliver2013tlsh}}
& 2.88 & 2.66 & 2.62 & 2.46 & 2.60 \\
\multicolumn{1}{c}{LZJD~\cite{raff2018lzjd}}
& 5.90 & 3.20 & 2.78 & 3.24 & 3.42 \\
\multicolumn{1}{c}{sdhash~\cite{roussev2010fingerprinting}}
& 33.28 & 11.08 & 3.34 & 7.54 & 8.28 \\
\midrule
\multicolumn{1}{c}{REM ($\tau=0.95$, Top3)}
& 330.54 & 91.02 & 2.44 & 25.20 & 28.46 \\
\multicolumn{1}{c}{REM ($\tau=0.91$, Top3)}
& 724.92 & 162.24 & 3.16 & 41.94 & 40.94 \\
\multicolumn{1}{c}{REM ($\tau=0.90$, Top7)}
& \textbf{979.32} & \textbf{207.72} & 3.62
& \textbf{52.94} & \textbf{49.84} \\
\bottomrule[1.1pt]
\end{tabular}%
}
\end{table*}

\begin{table*}[t]
\centering
\renewcommand{\arraystretch}{1.15}
\setlength{\tabcolsep}{5pt}
\caption{Robustness-design comparison under random scan-grid shifts.}
\label{tab:robustness_ablation}

\noindent\makebox[\linewidth][c]{%
\resizebox{0.9\linewidth}{!}{%
\begin{tabular}{cccccccc}
\toprule[1.1pt]
Configuration & Model Variant & AVHunter & DefenderCheck & HDD & HDD+ProbDD & HDD+CDD & Average \\
\midrule
\multirow{2}{*}{REM ($\tau=0.95$, Top3)}
& Full design
& 6,798.10 & 8,665.48 & 40,111.84 & 22,630.34 & 24,336.68 & \textbf{20,508.49} \\
& Without robustness
& 4,360.74 & 5,446.62 & 37,974.52 & 20,131.12 & 20,916.60 & 17,765.92 \\
\midrule
\multirow{2}{*}{REM ($\tau=0.91$, Top3)}
& Full design
& 14,802.56 & 15,982.94 & 41,770.28 & 28,426.94 & 29,616.06 & \textbf{26,119.76} \\
& Without robustness
& 9,442.38 & 10,720.34 & 42,428.80 & 27,858.90 & 28,645.10 & 23,819.10 \\
\midrule
\multirow{2}{*}{REM ($\tau=0.90$, Top7)}
& Full design
& 19,816.10 & 20,355.00 & 43,000.28 & 33,254.96 & 33,474.68 & \textbf{29,980.20} \\
& Without robustness
& 16,364.06 & 16,703.74 & 43,008.00 & 33,109.72 & 33,927.72 & 28,622.65 \\
\bottomrule[1.1pt]
\end{tabular}%
}%
}
\par
\end{table*}

\subsection{RQ3: Design Choices and Boundary Point Counts}
We examine the choice of reference-block length and scanning stride, the trade-off between matching configurations and additional false positives, the contribution of robustness design to probing costs, and the number of distinct BPs encountered during probing.

\subsubsection{Reference-Block Length and Scanning Stride}
\label{sec:reference_block_stride}
Figure~\ref{fig:rq3_block_stride_trends} presents the measured trade-offs, with REM ($\tau=0.91$, Top3; observed additional FPR $<0.1\%$) as the reference configuration. Increasing reference-block length from 512 to 1,024 bytes raises probing queries to approximately $1.78\times$ the original level. Further increasing it to 2,048 bytes yields only a 32.85\% query increase, while substantially increasing CPU scanning overhead and improving coverage by only 0.97 percentage points. We therefore choose 1,024-byte reference blocks and then examine scanning stride. Smaller strides increase CPU overhead without improving coverage, whereas larger strides reduce coverage. A 128-byte stride provides a favorable balance between protection, coverage, and scanning overhead, yielding a 1,152-byte scanning window.

\subsubsection{Matching Threshold and Block Selection}
\label{sec:matching_threshold_and_block_selection}
The similarity threshold $\tau$ and the maximum number of accepted blocks per source sample $k$ jointly control the trade-off between protection and additional false positives. Lowering $\tau$ or increasing $k$ generally increases probing costs but may also introduce more false positives. We therefore select REM configurations through a two-stage exploration. First, we explore thresholds from 0.85 to 0.95 across different Top-$k$ settings, identifying $(\tau,k)=(0.95,3)$ for the 0.05\% additional FPR budget (Figure~\ref{fig:configuration_selection}(a)). This exploration also identifies $\tau=0.90$ and $0.91$ as promising thresholds for balancing protection and false positives. We then explore Top1--Top10 at these two thresholds, selecting $(\tau,k)=(0.91,3)$ and $(0.90,7)$ for the 0.10\% and 0.20\% budgets, respectively (Figure~\ref{fig:configuration_selection}(b)). The exploration uses the same temporally later negative test partition reported in RQ2. The three selected configurations yield observed additional FPRs of 0.0254\%, 0.0891\%, and 0.1909\%, all below their corresponding budgets on this partition.


\begin{table*}[t]
\centering
\renewcommand{\arraystretch}{1.15}
\caption{Additional scanning latency and storage overhead of SKS relative to unmodified ClamAV.}
\label{tab:rq4_overhead}

\resizebox{0.9\linewidth}{!}{%
\begin{tabular}{ccccccc}
\toprule[1.1pt]
\shortstack{Method /\\Configuration} &
\shortstack{Additional CPU\\Latency (ms)} &
\shortstack{CPU\\Overhead (\%)} &
\shortstack{Additional GPU\\Latency (ms)} &
\shortstack{GPU\\Overhead (\%)} &
\shortstack{SS Library\\Storage (MiB)} &
\shortstack{Storage\\Increase (\%)} \\
\midrule
SimHash~\cite{charikar2002similarity} & 4.32 & 1.14 & 8.53 & 2.25 & 2.13 & 0.95 \\
ssdeep~\cite{kornblum2006identifying}  & 4.23 & 1.11 & --- & --- & 4.93 & 2.20 \\
TLSH~\cite{oliver2013tlsh}    & 4.05 & 1.07 & --- & --- & 4.91 & 2.19 \\
LZJD~\cite{raff2018lzjd}    & 8.16 & 2.15 & --- & --- & 130.27 & 58.18 \\
sdhash~\cite{roussev2010fingerprinting}  & 32.60 & 8.58 & --- & --- & 9.22 & 4.12 \\
\midrule
REM ($\tau=0.95$, Top3)
& 62.45 & 16.44 & 5.62 & 1.48 & 17.60 & 7.92 \\
REM ($\tau=0.91$, Top3)
& 64.65 & 17.02 & 2.54 & 0.67 & 17.60 & 7.92 \\
REM ($\tau=0.90$, Top7)
& 67.64 & 17.81 & 2.80 & 0.74 & 39.81 & 17.84 \\
\bottomrule[1.1pt]
\end{tabular}%
}

\smallskip
\begin{minipage}{\linewidth}
\footnotesize
\justifying
\noindent
Latency averages paired end-to-end measurements across the full dataset, excluding loading and file reads; overhead is normalized to the pooled 379.81\,ms ClamAV baseline. Storage increases are relative to the 223.91\,MiB frozen ClamAV database. SS storage includes matching representations and metadata, excluding raw blocks. REM uses FP16 storage and FP32 computation; its storage increase includes the 0.134\,MiB shared model.
\end{minipage}
\end{table*}

\begin{figure*}[t]
\centering
\includegraphics[width=0.9\linewidth]{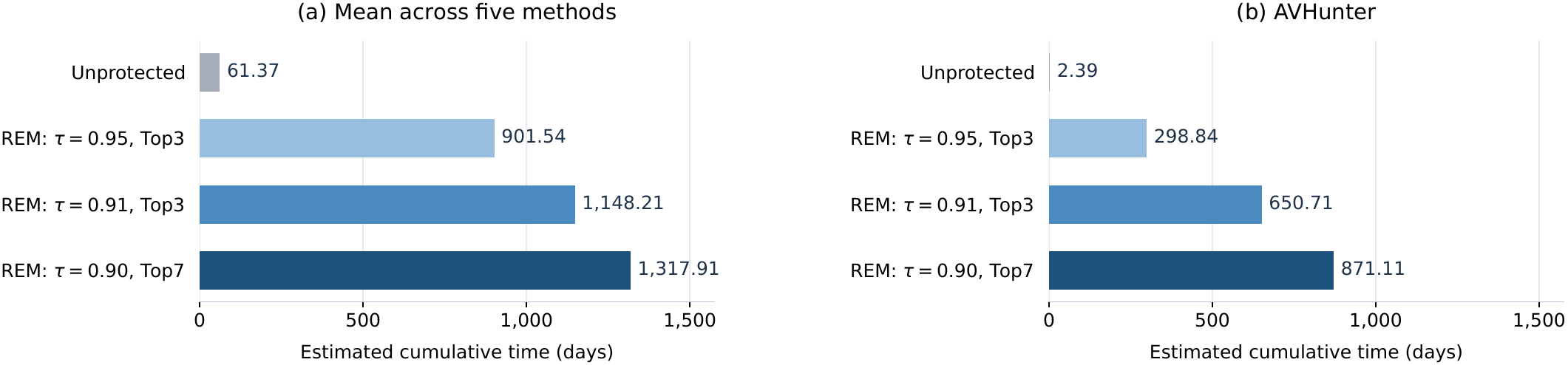}
\caption{Estimated serial probing time for 10,000 samples under random scan-grid shifts, comparing unprotected ClamAV with three REM configurations using mean queries across five probing methods and AVHunter alone.}
\label{fig:probing_time_10000}
\end{figure*}

\begin{table*}[t]
\centering
\renewcommand{\arraystretch}{1.1}
\caption{Effect of vector precision on training coverage,
additional test false positives, and SS storage.}
\label{tab:precision_comparison}

\resizebox{0.8\linewidth}{!}{%
\begin{tabular}{cccccc}
\toprule[1.1pt]
\shortstack{REM\\Configuration}  &
\shortstack{Vector\\Precision} &
\shortstack{Covered Training\\Positives} &
\shortstack{Training\\Coverage (\%)} &
\shortstack{Additional Test\\FPR (\%)} &
\shortstack{SS Library\\Storage (MiB)} \\
\midrule
\multirow{3}{*}{REM ($\tau=0.95$, Top3)}
& FP32 & 65,436 & 94.1878 & 0.0254 & 34.65 \\
& FP16 & 65,436 & 94.1878 & 0.0254 & 17.60 \\
& FP8  & 65,385 & 94.1143 & 0.0254 &  8.93 \\
\midrule
\multirow{3}{*}{REM ($\tau=0.91$, Top3)}
& FP32 & 66,508 & 95.7308 & 0.0891 & 34.65 \\
& FP16 & 66,508 & 95.7308 & 0.0891 & 17.60 \\
& FP8  & 66,503 & 95.7236 & 0.0891 &  8.93 \\
\midrule
\multirow{3}{*}{REM ($\tau=0.90$, Top7)}
& FP32 & 67,151 & 96.6563 & 0.1909 & 78.40 \\
& FP16 & 67,151 & 96.6563 & 0.1909 & 39.81 \\
& FP8  & 67,150 & 96.6549 & 0.1909 & 20.21 \\
\bottomrule[1.1pt]
\end{tabular}%
}

\smallskip
\begin{minipage}{\linewidth}
\footnotesize
Coverage is evaluated on 69,474 training positives, and additional
FPR on 7,859 test negatives. Encoding and similarity computation
use FP32 for all variants.
\end{minipage}
\end{table*}

\begin{figure*}[t]
\centering
\includegraphics[width=\linewidth]{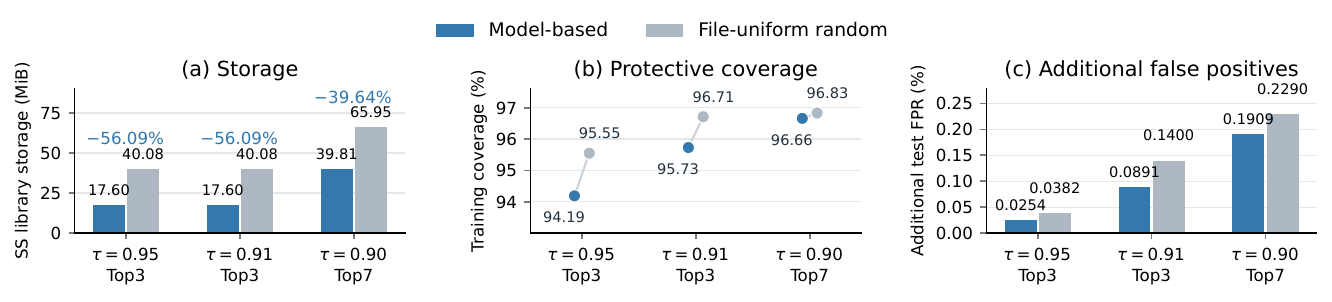}
\caption{Effect of model-based and file-uniform random block selection on (a) SS library storage, (b) training-sample coverage, and (c) additional test false-positive rates. Percentages above the bars in (a) indicate storage reductions achieved by model-based selection.}
\label{fig:model_random_selection}
\end{figure*}

\subsubsection{Ablation of Robustness Design}
\label{sec:robustness_ablation}

Table~\ref{tab:robustness_ablation} compares the full REM design with a variant that removes random-context and byte-perturbation training and skips incomplete tail windows, using the same 500 samples as RQ1 and fixed random scan-grid shifts. The full design increases average probing queries by 15.44\%, 9.66\%, and 4.74\% for $(\tau,k)=(0.95,3)$, $(0.91,3)$, and $(0.90,7)$, respectively. Gains are most pronounced for AVHunter and DefenderCheck, whereas HDD-based methods show smaller or mixed differences. These results support retaining the combined robustness design. Since the variants use separately trained models and their frozen SS libraries, the comparison reflects overall configuration differences rather than the isolated effect of training augmentation.

\subsubsection{Distinct Boundary Point Counts}
\label{sec:boundary_behavior}

Table~\ref{tab:rq3_boundary_points} reports the mean number of distinct BPs within the 50,000-query budget. Unprotected probing reports only 2.32--2.72 BPs, whereas REM increases BP counts for four of the five methods. The $(\tau=0.90,k=7)$ configuration reaches 979.32 BPs for AVHunter and 207.72 for DefenderCheck, exceeding all non-learning baselines. We hypothesize that the numerous BPs generated by learned SS matching, whose relation to byte-pattern rules is difficult to interpret, make PS recovery less likely. Even without overlapping SS regions, PS-related BPs may be difficult to distinguish from SS-induced ones, complicating analysis of genuine signature knowledge. HDD requires many queries per BP, so fewer reported BPs do not imply ineffective protection. More BPs generally accompany higher probing costs, with exceptions reflecting localization costs.

\begin{tcolorbox}[colback=gray!20,colframe=gray!20,boxrule=0pt,arc=0pt,left=6pt,right=6pt,top=6pt,bottom=6pt]
\textbf{Answer to RQ3.} The configuration analysis supports our parameter choices. Matching thresholds and block-retention limits support configurations with different additional-FPR budgets. Robustness design consistently increases average probing costs across the three REM configurations. REM also substantially increases reported BP counts for most probing methods, supporting the use of auxiliary matching to complicate PS-related feedback.
\end{tcolorbox}

\subsection{RQ4: Runtime and Storage Overhead}

\label{sec RQ4: Runtime and Storage Overhead}

We evaluate the scanning and storage overhead of SKS, focusing on how key design choices, including vector precision and block selection, affect SS library storage.

\subsubsection{Scanning and Storage Overhead}
Table~\ref{tab:rq4_overhead} reports the additional scanning latency and storage overhead of SKS. REM uses a learned encoder and a vector-based SS library, introducing 16.44\%--17.81\% CPU overhead. Its SS libraries and shared model add 7.92\%--17.84\% storage relative to the 223.91\,MiB ClamAV signature database. Given its consistent protection across probing methods and alignment settings, we consider this an acceptable trade-off. In GPU-equipped environments, additional scanning overhead is only 0.67\%--1.48\%. For latency-sensitive CPU-only deployments, non-learning methods remain alternatives. SimHash, for example, adds only 1.14\% CPU overhead and 0.95\% storage, although its protection weakens under scan-grid shifts.

Figure~\ref{fig:probing_time_10000} estimates cumulative serial probing time for 10,000 samples by multiplying the mean query counts under scan-grid shifts by the average ClamAV latency of 379.81\,ms. Using the mean across five methods, the estimate increases from 61.37 days without protection to 901.54--1,317.91 days with REM. Even for AVHunter, the most query-efficient evaluated method in these settings, it increases from 2.39 days to 298.84--871.11 days. These estimates illustrate how modest detection overhead can accompany substantially higher probing costs, making large-scale extraction of AV knowledge more difficult under practical time and resource constraints.

\subsubsection{Matching-Vector Precision} Table~\ref{tab:precision_comparison} compares SS vector storage precisions while keeping encoding and similarity computation in FP32. FP16 reduces library storage by approximately 49\% relative to FP32, with identical training coverage and additional test FPR across all three REM configurations. FP8 further reduces storage by approximately 49\% relative to FP16 and preserves the observed additional test FPR, but slightly decreases training coverage, with up to 51 fewer covered samples. We therefore use FP16 as the default storage format, substantially reducing storage without an observed loss in coverage or change in additional false positives.

\subsubsection{Effect of Model-Based Block Selection}

Figure~\ref{fig:model_random_selection} compares model-based selection with file-uniform random selection. All other construction, filtering, matching, and storage policies remain identical. Importance-guided selection reduces SS storage by 39.64\%--56.09\%, with only a 0.17--1.36 percentage-point decrease in training coverage. This suggests that the model favors blocks shared across samples, enabling a smaller SS library to achieve comparable protective coverage. It also yields lower observed additional test FPRs across all three configurations. These results support our choice of importance-guided selection, which substantially reduces storage requirements with only a small coverage loss while limiting additional false positives.

\begin{tcolorbox}[colback=gray!20,colframe=gray!20,boxrule=0pt,arc=0pt,left=6pt,right=6pt,top=6pt,bottom=6pt]
\textbf{Answer to RQ4.} SKS supports practical deployment with different resource constraints. REM achieves low additional scanning latency in GPU-equipped environments, while lightweight non-learning matchers offer alternatives for deployments with stringent detection-efficiency requirements. FP16 storage preserves the observed coverage and false-positive rates, and importance-guided block selection substantially reduces storage with only a small coverage loss while reducing additional false positives.
\end{tcolorbox}

\section{Related Work}
\label{sec:related_work}

\subsection{Signature-based Antivirus Detection}

Signature matching remains central to modern AV systems, while learning-based detection faces practical reliability challenges~\cite{botacin2022antiviruses,cavallaro2023machine}. Symantec, Trend Micro, and ESET retain signature and pattern matching~\cite{symantec_endpoint_security,trendmicro_apexone_patterns,eset_threatsense}; Kaspersky, FortiSandbox, and Bitdefender support YARA~\cite{kaspersky_yara_scan,fortinet_yara_rules,bitdefender2024yara}. ClamAV is integrated into Cisco Secure Endpoint on Linux, selected Barracuda appliances, and OPSWAT MetaDefender~\cite{cisco_secure_endpoint_clamav,barracuda_antivirus_technology,opswat2023metadefender}. Its supported signatures include file and section hashes, wildcard byte patterns, logical signatures, and YARA rules~\cite{clamav_signatures}. These rules encode accumulated expertise, but localized matching can expose detection-relevant content through scanning feedback.

\subsection{Extraction and Exploitation of Antivirus Knowledge}

Hamlen et al.~\cite{hamlen2009interface} investigate signature inference through AV interfaces, while Wressnegger et al.~\cite{wressnegger2017inferring} extract approximate signatures to induce false positives. DefenderCheck and ThreatCheck locate detection-triggering bytes through input splitting~\cite{defendercheck,threatcheck}, with delta debugging and HDD providing related reduction strategies~\cite{zeller2002simplifying,misherghi2006hdd}. Rigaki and Garcia~\cite{rigaki2023stealing} study surrogate-model extraction and evasion under low-false-positive conditions. AVHunter~\cite{yang2026breakingblackboxbytelevel} infers byte-level decision-critical regions to reproduce AV behavior and guide evasion. Bi-Iocane further manipulates AV-sensitive bytes to corrupt AV-generated labels and poison downstream ML detectors~\cite{yang2026weaponizinggroundtruthdata}. These studies motivate protecting detection knowledge against extraction, reuse, and exploitation.

\subsection{Approximate Matching and Knowledge Protection}

SimHash~\cite{charikar2002similarity}, ssdeep~\cite{kornblum2006identifying}, TLSH~\cite{oliver2013tlsh}, LZJD~\cite{raff2018lzjd}, and sdhash~\cite{roussev2010fingerprinting} use statistical features or compact digests to measure content similarity. Prior work emphasizes accuracy, robustness, and efficiency, with limited attention to protecting existing signatures. Filiol~\cite{filiol2006patterns} resists black-box analysis through Boolean detection patterns and combinatorial, probabilistic scanning, requiring changes to detection patterns and scanning logic. SKS preserves PS and their detection logic, integrating five tuned non-learning matchers as alternatives and baselines for auxiliary protection. REM explores learned matching to increase probing costs while limiting additional false positives.

\section{Discussion}
\subsection{Deployment Considerations and Adaptive Adversaries}

\paragraph{Why SS complement rather than replace PS} SS protect PS rather than improve detection generalization. The importance model uses file-level AV supervision, REM learns from reference blocks and augmented views, and filtering excludes candidates matching calibration negatives. SS thus inherit existing detection knowledge rather than independently validated knowledge of new threats. Their coverage concentrates on construction samples, whereas PS may detect other samples satisfying the same signature conditions. Replacing PS could therefore lose detections. SS matches may also fail while PS-related content remains intact, and do not inherently reproduce detection names or rule-specific conditions~\cite{clamav_signatures}. SKS therefore retains PS and uses SS for auxiliary protection.

\paragraph{When to use non-learning matchers} SimHash offers a lightweight SKS alternative, adding only 1.14\% CPU overhead and 0.95\% storage, with 98.52\% training coverage and an observed additional test FPR of 0.1272\%, below the 0.20\% budget. Although protection weakens under scan-grid shifts, it still substantially increases probing costs, supporting deployments requiring low overhead, broad known-sample coverage, and limited additional false positives.

\paragraph{Protection against informed adversaries} Adversaries may know the SKS design and matching procedure, but cannot directly access PS or SS libraries or modify scanner logic. Protection does not rely on algorithm secrecy, although informed adversaries may disrupt SS matches while preserving PS-related content. Scan-grid-shift experiments show stronger protection from REM than the evaluated non-learning alternatives, but other adaptive strategies require evaluation~\cite{tramer2020adaptive,feng2023stateful}. Our ClamAV-based implementation provides an optional feature that prioritizes blocks containing PS-matching regions for Top-$k$ selection after negative-sample filtering, offering a possible strategy against adaptive attacks. We encourage researchers to build on this implementation to further investigate defenses against such attacks~\cite{he2024malwaretotal,ling2024wolf}.

\subsection{Limitations and Future Work}

\paragraph{Scalability} Our evaluation covers 176,892 PE files, with latency measured across the full dataset. Larger sample collections may require larger SS libraries, increasing matching time and storage. Since REM represents SS as vectors, partition-based indexing~\cite{xu2023spfresh}, graph-based approximate retrieval~\cite{ootomo2024cagra}, and vector quantization~\cite{aguerrebere2023similarity,gao2024rabitq,gao2025practical} offer potential optimization directions. Approximate retrieval may miss SS matches and thereby alter protection effectiveness; future work must evaluate this trade-off to balance efficiency and protective coverage.

\paragraph{Additional False Positives} Even low additional false-positive rates remain a concern. Calibration and validation negatives may not adequately represent future inputs under distribution shifts~\cite{yang2021cade,chen2023continuous,li2025revisiting}. Expanding their scale and diversity, particularly with newer samples, may improve generalization and enable stronger protection under the same false-positive budget. This benefit must be balanced against possible SS coverage losses from more extensive filtering.

\paragraph{Protection Scope} SKS protects static detection signatures for Windows PE files, excluding dynamic behavioral monitoring and cloud-based analysis. Future work can extend the shielding principle to these mechanisms and to Android and Linux, adapting candidate selection and matching to their file structures and detection mechanisms.

\section{Conclusion}

We presented SKS, a design paradigm that protects existing antivirus signature knowledge through auxiliary SS matching. SKS combines AV-supervised block discovery, calibrated SS construction, and auxiliary matching to complicate signature probing while preserving original positive verdicts. Within this framework, we integrated five tuned non-learning matchers and introduced REM, a lightweight learned matcher trained with random context and byte perturbations. Evaluation of our open-source ClamAV-based prototype, which serves as a reference for knowledge protection in closed-source AV products, shows that SKS substantially increases probing costs while limiting additional false positives. REM provides stronger and more consistent protection under scan-grid shifts with low GPU scanning overhead, while non-learning alternatives support lightweight deployments. These results demonstrate the potential of auxiliary matching to protect existing detection knowledge. Future work will investigate broader adaptive attacks and improve scalability while maintaining protective coverage and low additional false-positive rates.

\section{Data Availability}

We provide a ClamAV-based SKS reference implementation for protecting signature knowledge in closed-source AV products, with SS libraries for the three evaluated configurations, at \url{https://anonymous.4open.science/r/SKS-CBF7/}. The repository includes model checkpoints, experiment configurations, evaluation scripts, reproduction instructions, and SHA-256 manifests for the 176,892-file dataset and 500-sample probing cohort. Researchers may request samples by providing their affiliation and research purpose, subject to data-sharing restrictions. Upon acceptance, these artifacts will be publicly released under the same access policy, with samples available by request.
%
%


\bibliographystyle{IEEEtran}
\bibliography{references}

\end{document}